\documentclass[]{aa}

\usepackage{graphicx}
\usepackage{txfonts}
\usepackage{hyperref}
\DeclareMathOperator{\sinc}{sinc}

\begin{document} 

   \title{Vector speckle grid: instantaneous incoherent speckle grid for high-precision astrometry and photometry in high-contrast imaging}
   
   \author{S.P. Bos}
   \institute{Leiden Observatory, Leiden University, P.O. Box 9513, 2300 RA Leiden, The Netherlands\\
                \email{stevenbos@strw.leidenuniv.nl}  
             }

   \date{Received {March 15, 2020}; accepted {May 4}, 2020}

 
  \abstract
   {Photometric and astrometric monitoring of directly imaged exoplanets will deliver unique insights into their rotational periods, the distribution of cloud structures, weather, and orbital parameters. 
    As the host star is occulted by the coronagraph, a speckle grid (SG) is introduced to serve as astrometric and photometric reference. 
    Speckle grids are implemented as diffractive pupil-plane optics that generate artificial speckles at known location and brightness.  
    Their performance is limited by the underlying speckle halo caused by {evolving uncorrected} wavefront errors.
    The speckle halo will interfere with the coherent SGs, {affecting} their photometric and astrometric precision.}
   {Our aim is to show that by imposing opposite amplitude {or} phase modulation on the opposite polarization states, a SG can be instantaneously incoherent with the underlying halo, greatly increasing the precision. 
    We refer to these as vector speckle grids (VSGs).}
   {We derive analytically the mechanism by which the incoherency arises and explore the performance gain in idealised simulations {under various atmospheric conditions}.}
   {We show that the VSG is completely incoherent for unpolarized light and that the fundamental limiting factor is the cross-talk between the speckles in the grid. 
   {In simulation, we find that for short-exposure images the VSG reaches a $\sim$0.3-0.8\% photometric error and $\sim$$3-10\cdot10^{-3}$ $\lambda/D$ astrometric error, which is a performance increase of a  factor $\sim$20 and $\sim$5, respectively.} 
   Furthermore, we outline how VSGs could be implemented using liquid-crystal technology to impose the geometric phase on the circular polarization states.}
   {The VSG is a promising new method for generating a photometric and astrometric reference SG that {has} a greatly increased astrometric and photometric precision.}


\titlerunning{Vector speckle grid: instantaneous incoherent speckle grid for high precision astrometry and photometry}
\maketitle
\section{Introduction}\label{sec:introduction}
Temporally monitoring directly imaged exoplanets will deliver unique insight into their rotational periods, and the distribution and dynamics of cloud structures \citep{kostov2012mapping}.
For example, HST observations showed that 2M1207b has photometric variations at the 0.78-1.36\% level, which allowed for the first measurement of the  rotation period of a directly imaged exoplanet \citep{zhou2016discovery}.
Furthermore, high precision astrometric monitoring of exoplanets will help determine their orbital dynamics.
This was demonstrated by \cite{wang2016orbit}, where {the authors showed} that $\beta$ Pic b does not transit its host star, but that its Hill sphere does. \\ 
\indent Ground-based high-contrast imaging (HCI) systems, such as SPHERE \citep{beuzit2019sphere}, GPI \citep{macintosh2014first}, and SCExAO \citep{jovanovic2015subaru}, deploy extreme adaptive optics (AO) systems to measure and correct for the turbulence in the Earth's atmosphere. 
The direct photometric and astrometric reference, that is the host star, is occulted by a coronagraph to reveal the faint companions. 
{This makes it hard to disentangle exoplanet brightness variations, due to their intrinsic rotation, from seeing and transmission changes in the Earth's atmosphere.}
To circumvent this problem, \cite{marois2006accurate} and \cite{sivaramakrishnan2006astrometry} simultaneously came up with diffractive methods to generate artificial speckle grids (SGs) that could serve as photometric and astrometric references. 
These are implemented as static masks {that introduce phase or amplitude modulations in the pupil plane, and are installed before the focal-plane mask of the  coronagraph.} 
{The artificial speckles are designed to fall on specific off-axis focal-plane positions and will therefore not be occulted by the coronagraph.} 
For example, GPI implemented a square grid that acts as an amplitude grating and reports a $\sim$7\% photometric stability \citep{wang2014gemini}, and SPHERE uses a static deformable mirror (DM) modulation with a $\sim$4\% photometric stability {(\citealt{langlois2013infrared}; \citealt{apai2016high}).} 
{The limiting factor of these SGs is their coherency with the time-varying speckle background, which}
{results in interference that dynamically distorts the shape and brightness of the SGs, which in turn ultimately limits their photometric and astrometric precision.} 
{The background speckles are for example generated by uncorrected wavefront errors due to fitting, bandwidth, or calibration errors in the AO system (\citealt{sivaramakrishnan2002speckle}; \cite{macintosh2005speckle}), or evolving non-common path errors \citep{soummer2007speckle}.}
{These background speckles have been found to decorrelate, that is, they become incoherent over timescales from seconds to minutes and hours  (\citealt{fitzgerald2006speckle}; \citealt{hinkley2007temporal};  \cite{martinez2012speckle}; \cite{martinez2013speckle}; \citealt{milli2016speckle}), and therefore affect the stability of the SGs during the entire observation.} \\
\indent \cite{jovanovic2015artificial} presented a method that circumvents this problem.  
Their solution is a high-speed, temporal modulation that switches (< 1 ms) the phase of the SG between zero and $\pi$ (e.g. by translating the mask). 
Due to the modulation, the interference averages out and the SG effectively becomes incoherent.
This has been implemented at SCExAO using DM modulation and is reported to increase the stability by a factor of between approximately two and three \citep{jovanovic2015artificial}.
However, a dynamic, high-speed component cannot always be integrated and is not always desired in a HCI system. 
For an implementation by DM modulation, the SG can only be placed within the control radius of the AO, and the incoherency relies on the quality of the DM calibration. \\
\indent Here, we propose the vector speckle grid (VSG).
This is a SG solution that {instantaneously} generates incoherency by imposing opposite amplitude {or} phase modulation on the opposite polarization states in the pupil plane.
{The opposite polarization states will both generate SGs at the same focal-plane positions, but with opposite phase.
Therefore, the two polarization states will interfere differently with the background speckle halo, but such that in total intensity the interference terms cancel.}
{The VSG} can be implemented as one static, liquid crystal optic that can be easily calibrated before observations. 
Furthermore, the speckles can be positioned anywhere in the focal plane and thus outside the scientifically interesting AO control region. \\
\indent In \autoref{sec:theory} we derive the theory behind the VSG.
In \autoref{sec:simulations} we perform idealised simulations to quantify the performance increase and investigate the effect of partially polarized light. 
In \autoref{sec:implementation} we discuss how the VSGs could be implemented. 
Lastly, in \autoref{sec:conclusions}, we discuss the results and present our conclusions.  
\section{Theory}\label{sec:theory} 
\begin{table*}
\caption{Variables presented in \autoref{sec:theory}.}
\label{tab:theory_variables}
\vspace{2.5mm}
\centering
\begin{tabular}{l|l}
\hline
\hline
Variable & Description \\ \hline
$\alpha$ & Real part of the aberration focal-plane electric field (\autoref{eq:aberration}).\\
$\beta$ & Imaginary part of the aberration focal-plane electric field (\autoref{eq:aberration}). \\
$\theta$ & Pupil-plane electric field phase.\\ 
$\Gamma$ & Real part of the speckle grid's focal-plane electric field (\autoref{eq:gamma}). \\
$\Delta$ & Real part of the central PSF's focal-plane electric field (\autoref{eq:delta}).\\
$\Omega$ & Imaginary part of the speckle grid's focal-plane electric field (\autoref{eq:omega}). \\
$\Lambda$ & Imaginary part of the central PSF's focal-plane electric field (\autoref{eq:lambda}) \\
$a$ & Amplitude of the speckle grid generating pupil-plane modulation (\autoref{eq:pupil_VPSG}). \\
$b$ & Spatial frequency of the speckle grid generating pupil-plane modulation(\autoref{eq:pupil_VPSG}).\\
$f_x$ & Focal-plane coordinate.\\ 
$p$ & Degree of polarization.\\ 
$x$ & Pupil-plane coordinate.\\
$A$ & Pupil-plane electric field amplitude.\\
$E_f$ & Focal-plane electric field \\ 
$E_p$ & Pupil-plane electric field.\\ 
$\mathcal{F}\{\cdot\}$ & Fourier transform operator.\\ 
$I_f$ & Focal-plane intensity of the PSF.\\
$I_s$ & Relative intensity of the speckle grid.\\
\hline
\end{tabular}
\end{table*}
\subsection{Vector phase speckle grid}\label{sec:VPSG}
Here we derive how the incoherency of VSGs arises, and focus on the vector phase speckle grid (VPSG) first. 
The derivation of the vector amplitude speckle grid (VASG) is presented in \autoref{sec:VASG}.
{All variables used in this section are also defined in \autoref{tab:theory_variables}.}
We assume that the stellar point spread function (PSF) is only aberrated by phase aberrations for simplicity.
However, VSGs are still incoherent when there are also amplitude aberrations present.   
Here, we assume a one-dimensional pupil-plane electric field $E_p$(x):
\begin{equation}\label{eq:Epup}
E_p(x) = A(x) e^{i \theta(x)}, 
\end{equation} 
with $A(x)$ being the amplitude of the electric field, which is described as the rectangular function, $\theta(x)$ the phase aberration distorting the PSF, {and $i=\sqrt{-1}$ the unit imaginary number}.
The coordinate $x$ denotes the position in the pupil and is omitted from here on.  
We describe starlight with a degree of polarization $p$, as two orthogonal polarization states (either linear or circular) using Jones calculus:
\begin{equation}\label{eq:polarization}
E_1 = \frac{1}{\sqrt{2}}
\begin{pmatrix}
\sqrt{1+p} \\
0
\end{pmatrix}, \ \ E_2 = \frac{1}{\sqrt{2}}
\begin{pmatrix}
0 \\
\sqrt{1-p}
\end{pmatrix}.
\end{equation}
The VPSG is implemented by a cosine wave (with spatial frequency $b$) on the pupil phase, with opposite amplitude $a$ for the two opposite polarization states. 
{As we see below in the derivation, $b$ determines the focal-plane position of the artificial speckles and $a$ their relative brightness to the PSF core.}
The opposite amplitude eventually leads to the incoherency of the VSG.
\begin{equation}\label{eq:pupil_VPSG}
E_p = \frac{A e^{i \theta}} {\sqrt{2}} 
\begin{pmatrix}
\sqrt{1+p} \cdot e^{ai \cos(2 \pi b x) }\\ 
\sqrt{1-p} \cdot e^{-ai \cos(2 \pi b x)}
\end{pmatrix} 
.\end{equation}
We assume the Fraunhofer approximation and calculate the focal-plane electric field $E_f$ by taking the Fourier transform ($\mathcal{F}\{\cdot\}$) of $E_p$:
\begin{align}\label{eq:focal_VPSG}
E_f &= \mathcal{F} \{ E_p\} (f_x) \\
      &= \frac{\mathcal{F} \{ A\} * \mathcal{F} \{  e^{i \theta}\}}{\sqrt{2}} *  
      \begin{pmatrix}
\sqrt{1+p} \cdot \mathcal{F}\{e^{ai \cos(2 \pi b x)} \} \\ 
\sqrt{1-p} \cdot \mathcal{F}\{e^{-ai \cos(2 \pi b x)} \}
\end{pmatrix} ,
\end{align}
where {$*$ is the convolution operator, and} $f_x$ the coordinate in the focal plane, which is omitted as well.
{As we chose $A(x)$ to be the rectangular function in \autoref{eq:Epup}, its Fourier transform is}:
\begin{align}
\mathcal{F}\{ A \} (f_x) &= \sin(f_x) / f_x \label{eq:sinc_def} \\
                                   &= \sinc(f_x).
\end{align}
We do not explicitly calculate $\mathcal{F}\{e^{i \theta} \}$ and assume that:
\begin{equation}\label{eq:aberration}
\mathcal{F}\{e^{i \theta} \}  = \alpha + i \beta
.\end{equation}
Writing $e^{\pm ai \cos(2 \pi b x)}$ as a series expansion, we find that: 
\begin{equation}\label{eq:Ef_VPSG}
\begin{aligned}
E_f &= \frac{\sinc(f_x) * (\alpha + i \beta)}{\sqrt{2}} * \\
        &\qquad \begin{pmatrix}
\sqrt{1+p} [1 +  \sum_{n=1}^{\infty} \frac{(i)^n a^n}{n!} \mathcal{F}\{ \cos^n(2 \pi b x) \} ]\\ 
\sqrt{1-p}  [1 +  \sum_{n=1}^{\infty} \frac{(i)^n (-a)^n}{n!} \mathcal{F}\{ \cos^n(2 \pi b x) \} ]
\end{pmatrix}.
\end{aligned}
\end{equation}
\autoref{eq:Ef_VPSG} shows that the VPSG creates an infinite number of speckles with decreasing brightness. 
For now, we assume that $a \ll 1$ radian and expand \autoref{eq:Ef_VPSG} to first order ($n=1$). 
Working out the terms in \autoref{eq:Ef_VPSG}, we find:
\begin{equation}\label{eq:Efoc}
\begin{aligned}
E_f &=
\frac{\alpha + i \beta}{\sqrt{2}} * \\
&\qquad
\begin{pmatrix}
  \sqrt{1+p} [ \sinc(f_x) + \frac{ai}{2} (\sinc(f_x -b) + \sinc(f_x + b))] \\
   \sqrt{1-p}[ \sinc(f_x) - \frac{ai}{2} (\sinc(f_x -b) + \sinc(f_x + b))]
\end{pmatrix}.
\end{aligned}
\end{equation}
Rearranging in the real and imaginary terms, and computing the focal-plane intensity ($I_f = E_f \cdot E_f^*$) results in:
\begin{equation}\label{eq:focal_intensity}
I_f = \underbrace{\Delta^2 + \Lambda^2}_{\text{PSF}} + \underbrace{\frac{a^2}{4}(\Gamma^2 + \Omega^2)}_{\text{speckle grid}} + \underbrace{a p (\Delta \Gamma + \Lambda \Omega).}_{\text{cross-talk of PSF with speckle grid}}
\end{equation}
Greek symbols are used here to simplify the notation and denote the following terms: 
\begin{align}
\Delta     &=  \sinc(f_x) * \alpha \label{eq:delta} \\
\Gamma &= [\sinc(f_x -b) + \sinc(f_x + b)] * \beta \label{eq:gamma} \\
\Lambda &= \sinc(f_x) * \beta \label{eq:lambda} \\
 \Omega  &= [\sinc(f_x -b) + \sinc(f_x + b)] * \alpha. \label{eq:omega}
\end{align}
{\autoref{eq:focal_intensity} shows that the focal-plane intensity can be divided into three terms: the stellar PSF, the speckle grid, and the cross-talk between the speckle grid and the PSF.}
{We find} that the relative intensity of the speckle grid is given by $I_{s} = a^2 / 4$. 
{The performance of regular SGs is limited by the cross-talk term.}
{For the VPSG, when $p=0$ (i.e. with unpolarized light), the cross-talk term is eliminated} and {\autoref{eq:focal_intensity}} reduces to:
\begin{equation}\label{eq:focal_intensity_nocross}
I_f = \Delta^2 + \Lambda^2 + \frac{a^2}{4}(\Gamma^2 + \Omega^2),
\end{equation}
effectively making the VPSG incoherent with the PSF. \\
\indent The remaining cross-talk that degrades the photometric and astrometric performance is the interference between the speckles themselves:
\begin{equation}\label{eq:speckle_crosstalk}
\begin{aligned}
\Gamma^2 + \Omega^2 = &\underbrace{(\sinc(f_x -b) * \alpha)^2 + (\sinc(f_x -b) * \beta)^2}_{\text{speckle 1}} + \\
                                        &\underbrace{(\sinc(f_x +b) * \alpha)^2 + (\sinc(f_x +b) * \beta)^2}_{\text{speckle 2}} + \\
                                        & \underbrace{2 (\sinc(f_x - b) * \alpha \cdot \sinc(f_x + b) * \alpha}_{\text{cross-talk between speckles}}  + \\
                                        & \underbrace{\sinc(f_x - b) * \beta \cdot \sinc(f_x + b) * \beta)).}_{\text{cross-talk between speckles}}
\end{aligned}
\end{equation}
{We have not found a method to mitigate this effect, and consider this cross-talk to be the fundamental limiting factor of the VSG.}
Its effect can be reduced by minimizing the number of speckles in the VSG and increasing their separation. 
{This can be understood as follows: the distortion of an artificial speckle is determined by the relative strength of the  combined electric field of the other artificial speckles (the distorting electric field) at its location, relative to its own electric field strength.  
If the distorting electric field becomes stronger, the cross-talk terms in \autoref{eq:speckle_crosstalk} become more important and the artificial speckle is more distorted.
If there are fewer artificial speckles in the VSG, the distorting electric field becomes weaker. 
Furthermore, as the electric field an artificial speckles scales with $f_x^{-1}$ (\autoref{eq:sinc_def}), placing the artificial speckles further apart also reduces the distorting electric field. }\\
\indent We expanded \autoref{eq:Ef_VPSG} to first order and ignored higher order terms; we discuss their effects here. 
The higher order terms can be divided into two groups: the odd orders ($n=$ odd) and the even orders ($n=$ even). 
The amplitude of the higher order terms is given by $(a)^n$.
For the odd orders, $n$ is odd, and therefore, when $a$ flips its sign ($a\rightarrow-a$), the higher orders also have a sign flip. 
Which means that all the odd orders become incoherent as the cross-talk term between the PSF and the speckle grid cancels when $p=0$. 
For the even orders ($n=$ even), a sign flip of $a$ does not result in a sign flip of $(a)^n$.
This means that none of the even orders are incoherent as the cross-talk term does not cancel. 
As the even orders fall at other spatial locations and are much fainter than the first order speckles, the impact of the coherent even orders is minimal. 

\subsection{Vector amplitude speckle grid}\label{sec:VASG}
Here we derive how the incoherency of the VASG arises. 
This derivation is very similar to what is presented in \autoref{sec:VPSG} and starts with the same assumptions.
The VASG is implemented by a sine wave (with spatial frequency $b$) on the pupil amplitude, with opposite amplitude $a$ for the two opposite polarization states:
\begin{equation}
E_p = \frac{A e^{i \theta}} {\sqrt{2}} 
\begin{pmatrix}
\sqrt{1+p} [ 1 + a \sin(2 \pi b x)]  \\ 
\sqrt{1-p} [1 -a \sin(2 \pi b x)]
\end{pmatrix} 
.\end{equation}
We calculate the focal-plane electric field $E_f$ by taking the Fourier transform ($\mathcal{F}\{\cdot\}$) of $E_p$:
\begin{align}
E_f &= \mathcal{F}\{E_p \} \\ 
       &= \frac{\mathcal{F}\{A \} * \mathcal{F}\{e^{i \theta} \}} {\sqrt{2}} *  
      \begin{pmatrix}
      \sqrt{1+p} [1 + a \mathcal{F}\{\sin(2 \pi b x)\}] \\ 
      \sqrt{1-p} [1 - a \mathcal{F}\{\sin(2 \pi b x)\}]
      \end{pmatrix} \label{eq:Efoc_in_Epup_VASG}.
\end{align}
Using \autoref{eq:aberration} and working out the Fourier transforms of the terms in \autoref{eq:Efoc_in_Epup_VASG}, we find:
\begin{equation}
\begin{aligned}
E_f &=
\frac{\alpha + i \beta}{\sqrt{2}} * \\
  &\begin{pmatrix}
  \sqrt{1+p} [ \sinc(f_x) + \frac{ai}{2} (\sinc(f_x -b) - \sinc(f_x + b))]\\
  \sqrt{1-p}[ \sinc(f_x) - \frac{ai}{2} (\sinc(f_x -b) - \sinc(f_x + b))]
\end{pmatrix}.
\end{aligned}
\end{equation}
Rearranging this in its real and imaginary terms, and computing the focal-plane intensity ($I_f = E_f \cdot E_f^*$) results in:
\begin{equation}\label{eq:focal_intensity_VASG}
I_f = \underbrace{\Delta^2 + \Lambda^2}_{\text{PSF}} + \underbrace{\frac{a^2}{4}(\Gamma^2 + \Omega^2)}_{\text{speckle grid}} + \underbrace{a p (\Delta \Gamma + \Lambda \Omega).}_{\text{cross-talk of PSF with speckle grid}}
\end{equation}
{As in \autoref{sec:VPSG}, the Greek symbols denote the following terms:} 
\begin{align}
\Delta &=  \sinc(f_x) * \alpha\\
\Gamma &= [\sinc(f_x -b) - \sinc(f_x + b)] * \beta \\
\Lambda &= \sinc(f_x) * \beta \\ 
\Omega &= [\sinc(f_x -b) - \sinc(f_x + b)] * \alpha
\end{align}
In \autoref{eq:focal_intensity_VASG} we find again that the relative intensity of the speckle grid is given by $I_{s} = a^2 / 4$, and that the VASG becomes incoherent when $p=0$. 
As with the VPSG, the remaining cross-talk that degrades the photometric {and} astrometric performance is the interference between the speckles themselves. 
We also note that this implementation {with a sine wave modulation} of the VASG does not generate any higher order speckles.
{A VASG implementation comparable to the GPI amplitude grating \citep{wang2014gemini} will introduce higher order speckles in a  similar manner to the VPSG.}

\section{Simulations} \label{sec:simulations}
\subsection{Performance quantification}
\begin{table}
\caption{Parameters in the simulations {presented in \autoref{sec:simulations}.}}
\label{tab:simulation_parameters}
\vspace{2.5mm}
\centering
\begin{tabular}{l|l}
\hline
\hline
Parameter & Value\\ \hline
{Outer scale} & {30 m at 500 nm} \\ 
{Seeing} & {[0.6", 1", 1.4"] at 500 nm} \\ 
Wind speed & {[4.4, 8.8, 13.2] m/s} \\
 & \\
{Wavefront sensor} & {Noiseless} \\
{Frame rate}  & {2 kHz} \\

{Deformable mirror} & {40 $\times$ 40 actuators} \\
Lag of AO & 3 {frames}\\
Loop duration & 1 {s}\\
 & \\
Telescope diameter & 8 {m} \\
Wavelength & {1220 nm} \\
Bandwidth & Monochromatic \\
 & \\
{Coronagraph} & {Vector Vortex Coronagraph} \\
{Lyot stop diameter} & {0.95 $\cdot$ Telescope diameter} \\
 & \\
Speckle intensity & {$5\cdot10^{-3}$} \\ 
Speckle positions & {[$\pm$25 $\lambda / D$, $\pm$25 $\lambda / D$]} \\
\hline
\end{tabular}
\end{table}
\begin{figure*}
\centering
\includegraphics[width=17cm]{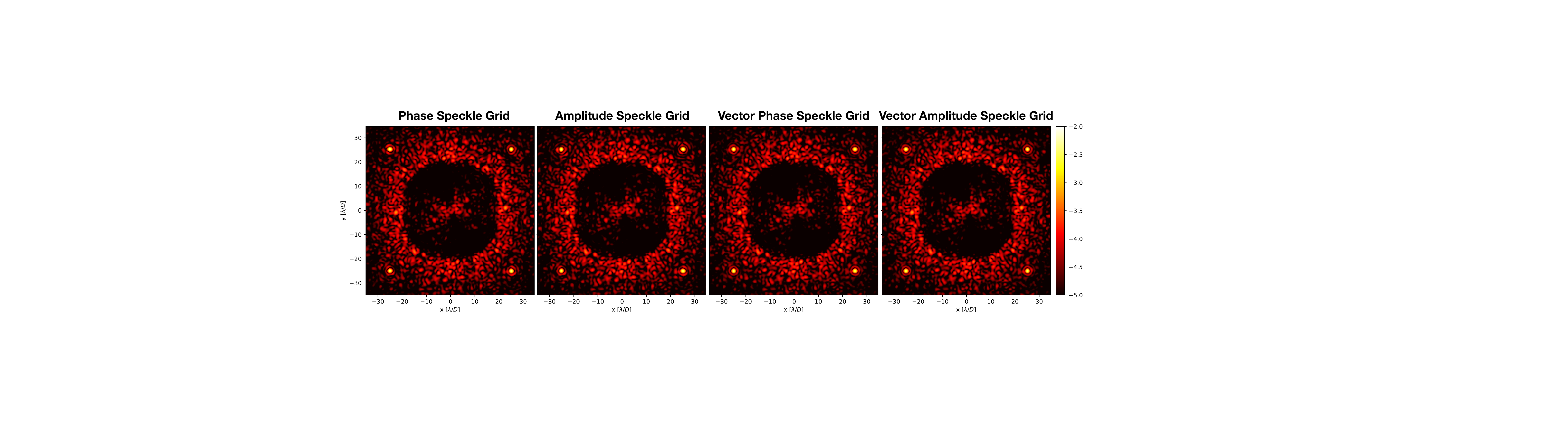}
\caption{
{Short-exposure images at a random iteration in the simulation with the medium atmospheric conditions (seeing of 1" and wind speed of 8.8 m/s).
The dark hole in the center of the images is generated by the AO system and the Vector Vortex Coronagraph.
The reference speckles are located in the corners of the image at [$\pm$25 $\lambda/D$, $\pm$25 $\lambda/D$].
We note that the reference speckles generated by the {VPSG and VASG} are significantly less distorted and more similar to each other compared to the Phase and Amplitude Speckle Grids.
The images show intensity normalized to the maximum of the star in logarithmic scale, and share the same color bar (shown at the right).}
}
\label{fig:single_frame}
\end{figure*}
\begin{figure}
\centering
\includegraphics[width=\hsize]{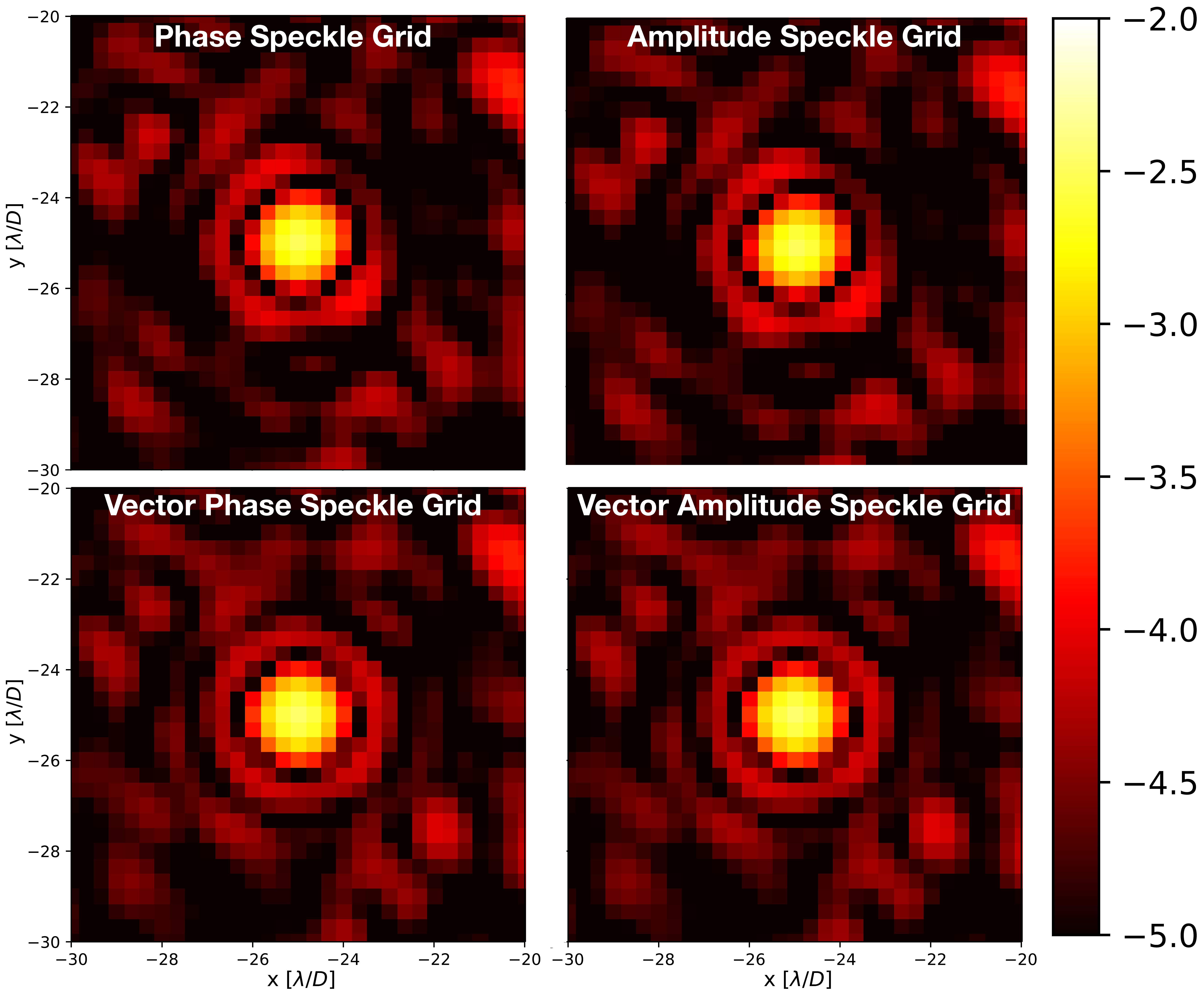}
\caption{ 
{Zoom onto the reference speckles in the lower left corner of the panels in \autoref{fig:single_frame}.
The VSGs are clearly less distorted compared to the regular SGs.
The images show intensity normalized to the maximum of the star in logarithmic scale, and share the same color bar (shown at the right).}
}
\label{fig:single_frame_ref_sub}
\end{figure}
\begin{figure*}
\centering
\includegraphics[width=17cm]{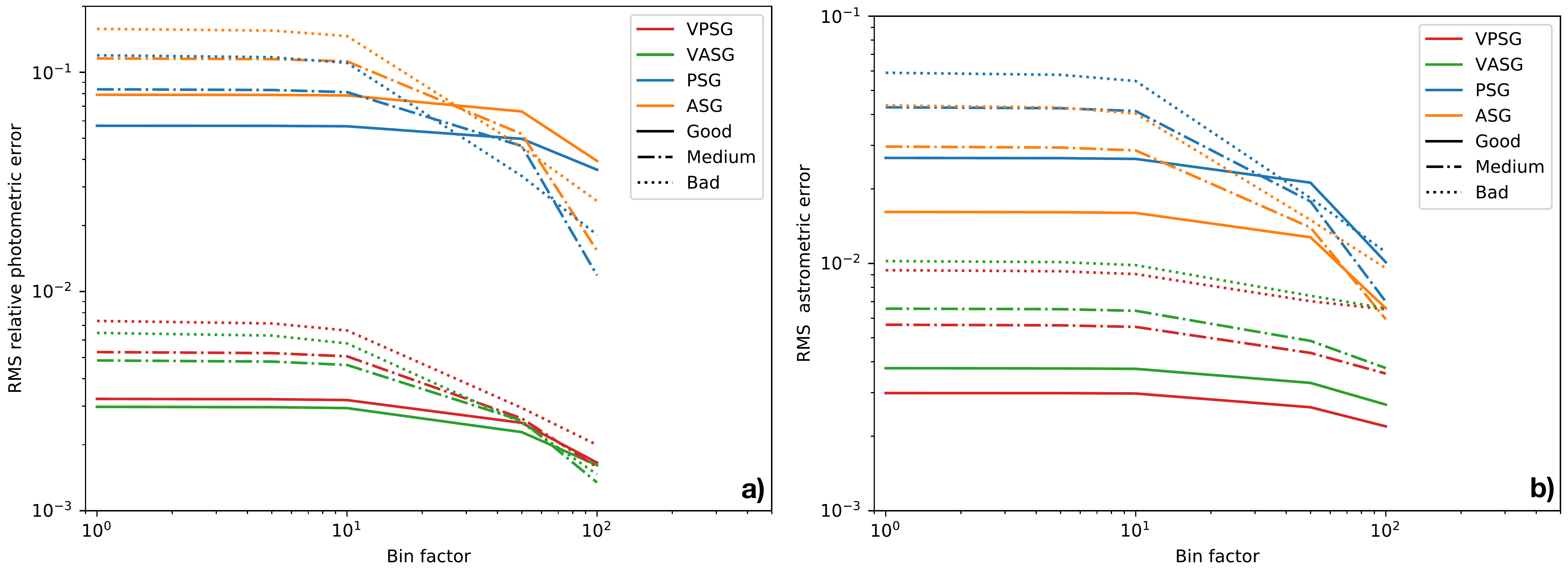}
\caption{Photometric and astrometric performance of the SGs as a function of the number of frames averaged{, and for various atmospheric conditions (\autoref{tab:simulation_parameters})}. (a) The rms relative photometric error. (b) The rms astrometric error.}
\label{fig:comparison_SGs}
\end{figure*}
To validate the VSG concept and quantify the performance increase {that VSGs could bring} compared to regular SGs, we performed {idealised numerical simulations}.  
{These} are performed in Python using the HCIPy package\footnote{\url{https://hcipy.org}} \citep{por2018hcipy}, which supports polarization propagation with Jones calculus necessary for simulating VSGs. 
{It is notoriously hard to realistically simulate high-contrast imaging observations as not all speckle noise sources are well understood \citep{guyon2019technology}, thus making it hard to predict the on-sky performance of the VSG.}
{Therefore, we decided to limit the scope of the simulations.}
{We simulated a general HCI instrument with coronagraph and AO system at an 8 m class telescope with a clear aperture under various atmospheric conditions, and did not include any other noise sources (e.g. detector and photon noise, evolving non-common path aberrations).}
An overview of the simulation parameters is shown in \autoref{tab:simulation_parameters}.
{We only considered monochromatic images as we leave broadband effects for future work.}
{The images are recorded at a wavelength of 1220 nm, which is at the centre of J-band, a scientifically interesting band for photometric variations of exoplanets \citep{kostov2012mapping}.}
{We considered three cases of atmospheric conditions, under which the current generation of HCI instruments regularly operate:}
\begin{enumerate}
\item {Good conditions with a seeing of 0.6" and wind speed of 4.4 m/s.}
\item {Medium conditions with a seeing of 1" and wind speed of 8.8 m/s.}
\item {Poor conditions with a seeing of 1.4" and wind speed of 13.2 m/s.}
\end{enumerate}
{These atmospheric conditions were simulated as} an evolving turbulent wavefront assuming the "Frozen Flow'' approximation with a von-karman power spectrum.
{The AO system that measures and corrects the atmospheric conditions consists of a noiseless wavefront sensor, and a deformable mirror with 40 $\times$ 40 actuators in a rectangular grid.}
{The AO system runs at 2 kHz and has a three-frame lag between measurement and correction.}  
{The root mean square (rms) residual wavefront error after the AO system is respectively 44 nm, 70 nm, and 95 nm for the good, medium, and poor atmospheric conditions.}
Following the Mar\'{e}chal approximation \citep{roberts2004really}, these residual wavefront errors correspond to Strehl ratios of 95\%, 88\%, and 79\%, respectively (calculated at $\lambda = 1220$ nm).
{With this setup we only consider the speckle noise from the free atmosphere.}
{As the decorrelation timescale for such speckles is on the order of $\sim$1 second \citep{macintosh2005speckle}, we limited the duration of the simulation to 1 second.}
{For longer simulations, the background speckles would effectively become incoherent with the SGs.}
{Focal-plane images were recorded at 2 kHz.}
{The coronagraph in this setup is the Vector Vortex Coronagraph (VVC; \citealt{mawet2005annular}) with an accompanying Lyot stop with a 95\% diameter.}
{The VVC is a focal-plane mask that removes starlight and operates on the vector state of light.}
{The reason for implementing the VVC in this simulation is twofold: first, for a clear aperture, the performance of the VVC is close to that of the perfect coronagraph \citep{cavarroc2006fundamental}, and second, as the VVC also operates on the vector state of light, we show that the performance of the VSG will not be affected by such optics.}
{The SGs are placed at $[\pm 25 \lambda / D, \pm 25 \lambda / D]$, which is well beyond the control radius of the AO system.} 
{The intensities of the SGs relative to the PSF core are $5 \cdot 10^{-3}$, making them $\sim$150, 63, and 35 times brighter than the background speckle halo for the good, medium, and poor atmospheric conditions, respectively}. 
We compare the phase speckle grid (PSG) and the amplitude speckle grid (ASG) to the VPSG and the VASG. 
We did not include any temporally modulated SGs because, when assuming an instantaneous modulation, their performance is equal to the {VSG}. 
{The VPSG and VASG are implemented on the circular polarization state.}
The simulations {were} performed with the same wavefronts for all the SGs. 
{We also performed the same simulations without a SG for accurate background estimation.} \\
\indent In \autoref{fig:single_frame} we show images of the SGs at a random iteration in the simulation {for medium atmospheric conditions}. 
{The coronagraph and AO system removed the central stellar light and generated the dark hole, which is clearly visible.} 
{The idealistic AO system gives an optimistic contrast in the dark hole.}
{Outside of the dark hole and control region, the speckle background is clearly visible.}
{This background is generated by residual wavefront errors and evolves during the simulation.} 
{It will strongly} interfere with the SGs, affecting their photometric and astrometric accuracy. 
{The reference speckles of the SG are located in the corners of the images ($\pm$25 $\lambda/D$, $\pm$25 $\lambda/D$).}
{The VPSG and VASG are significantly less distorted and more similar to each other compared to the PSG and ASG.}
{This shows the effect of the  incoherency of the VSG.}
Zooming in on the {lower left reference} speckles {as shown in \autoref{fig:single_frame_ref_sub} demonstrates this as well}. \\ 
\indent To quantify the performance increase offered by the VSGs, we calculate the rms photometric error and the rms astrometric error. 
These are calculated on the individual frames, and images that are averages of 5, 10, 50, and 100 frames to simulate longer exposure times. 
The photometric performance is calculated by carrying out the following steps: 
\begin{enumerate}
\item {Measure the photometry of the SG with an aperture with a diameter of $2.44$ $\lambda/D$}.
\item {Measure the background by aperture photometry in the simulation without SG at the same positions}.
\item Subtract the background estimate from the SG photometry.
\item Remove the general photometric fluctuations by dividing the SG photometry by the {mean photometry of the four speckles.}
\item Calculate the rms photometry error per speckle.
\item Determine the final rms photometric error by {calculating the mean of the four speckles individual rms photometric errors.}
\end{enumerate}
The astrometric performance is calculated by carrying out the following steps: 
\begin{enumerate}
\item Measure the position of the individual speckles by cross-correlation with {an unaberrated PSF}. 
\item Calculate the distance between the speckles. 
\item Calculate the rms of these distances over all images. 
\item Calculate the mean rms astrometric error over all the distances, which gives the final rms astrometric error. 
\end{enumerate}
In \autoref{fig:comparison_SGs} we plot the photometric and astrometric performance of the SGs as a function of the number of averaged frames {(bin factor) and for the different atmospheric conditions}. 
\autoref{fig:comparison_SGs} a shows that the VASG and the VPSG outperform the ASG and PSG in photometric error by a factor of {$\sim$10-20 (depending on the bin factor and atmospheric conditions)}. 
The VASG and VPSG {reach a $\sim$0.3-0.8\% photometric error for individual frames, which drops to $\sim$0.2\% when the bin factor} increases. 
The ASG and PSG on the other hand start at $\sim$6-15\% photometric error and decrease to $\sim$1.5-4\%.
\autoref{fig:comparison_SGs} b {also shows a }performance increase for the astrometric performance. 
The VASG and VPSG improve the astrometric error by a factor of $\sim$3-5 with respect to the ASG and PSG.  
At a bin factor of one, the VSGs reach an astrometric error of $\sim$$3-10\cdot10^{-3}$ $\lambda/D$, {and slightly improve} for a bin factor of 100.
{The ASG and PSG start at  $\sim$$1.5-6\cdot10^{-2}$ $\lambda/D$ and improve to $\sim$$7-16\cdot10^{-3}$ $\lambda/D$.} 
These results clearly demonstrate that VSGs greatly improve the photometric and astrometric precision with respect to their non-vector counterparts. \\
{For poorer seeing conditions, the performance of all SGs decreases. For the ASG and PSG, this is due to the increased speckle background halo that interferes with the SG (\autoref{eq:focal_intensity}), while for the VSGs this is due to the increased cross-talk between the reference speckles (\autoref{eq:speckle_crosstalk}).}
{When the wind speed increases, the performance of the SGs increases more rapidly with bin factor.}
{This is because the decorrelation timescale of background speckles scales with the inverse of the windspeed \citep{macintosh2005speckle}.}
{Therefore, for higher wind speeds, the interference between the background speckles and SGs will decrease with increasing bin factor, improving their performance.}

\subsection{Degree of polarization effects}\label{sec:DoP}
\begin{figure*}
\centering
\includegraphics[width=17cm]{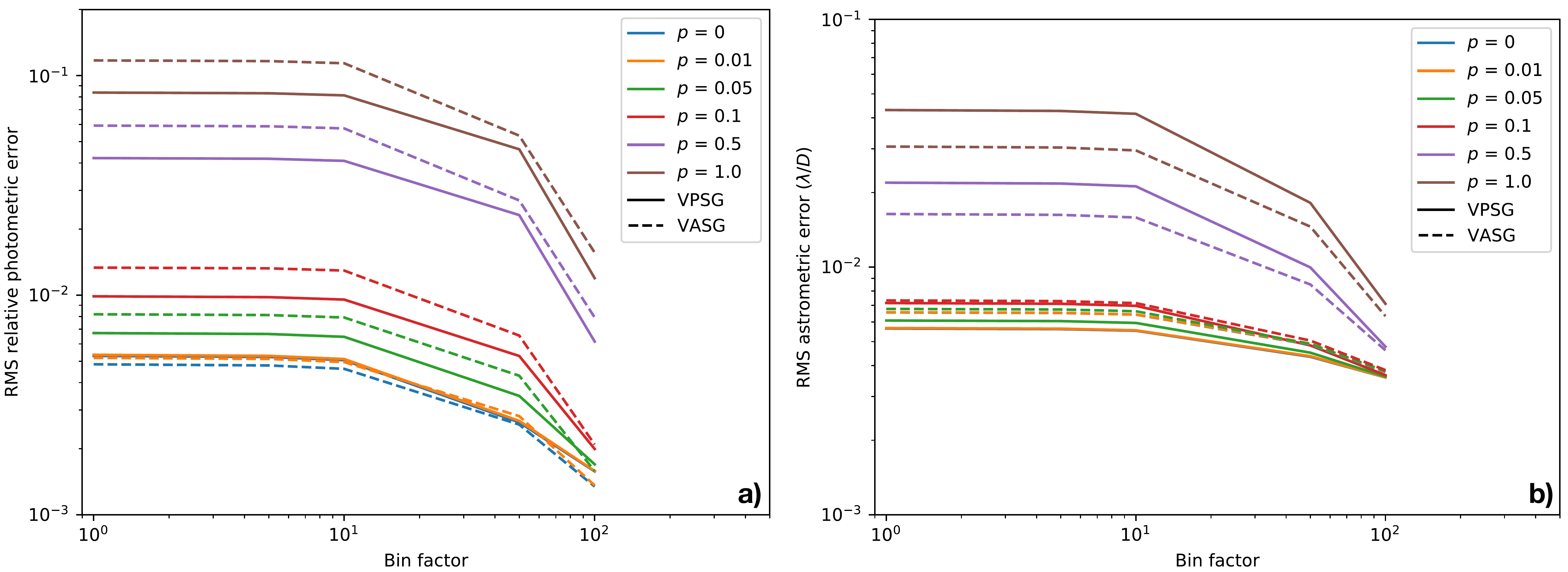}
\caption{Photometric and astrometric performance of the  VPSG and VASG as a function of the number of frames averaged and various degrees of polarization {($p$) under medium atmospheric conditions (\autoref{tab:simulation_parameters})}. (a) The rms relative photometric error. (b) The rms astrometric error.}
\label{fig:dop_effects}
\end{figure*}
As discussed in \autoref{sec:theory}, the incoherency of the VSGs depends on the degree of polarization {($p$; \autoref{eq:polarization})} of the light passing through the VSG (specifically the polarization state on which the VSG operates). 
As starlight  is generally unpolarized to a very high degree (e.g. the integrated $p$ of the Sun is $<10^{-6}$; \citealt{kemp1987optical}), we are mainly concerned with polarization introduced by the telescope and instrument. 
{For VLT/SPHERE, the linear $p$} has been measured to be on the order of a few percent \citep{van2020polarimetric} and the circular $p$ is expected to be even lower.
To study the effect of $p$, we repeat the simulations of \autoref{fig:comparison_SGs} {with the medium atmospheric conditions} for the VSGs but with increasing levels of $p$.
The simulations are performed with the same wavefronts as in \autoref{fig:comparison_SGs}. 
Therefore, {for $p=0$, we} expect exactly the same results, while {for $p=1$ the} performance of the VSGs should reduce to that of the ASG and PSG. 
\autoref{fig:dop_effects} (a) shows the photometric performance and \autoref{fig:dop_effects} (b) the astrometric performance. 
Both figures indeed show that the performance of the VASG and VPSG degrade to that of the ASG and PSG when $p = 1$, and are optimal for $p = 0$. 
They also show that up to a $p$ of 0.05  there is barely a performance degradation. 
The performance degrades close to linearly as a function of $p$, which is what we expect from \autoref{eq:focal_intensity}.

\section{Implementation of vector speckle grid}\label{sec:implementation}
Now that we have demonstrated that VSGs can drastically improve the photometric and astrometric performance, we discuss how VSGs can be implemented in a HCI system. 
We focus on the VPSG as we have not found a straightforward implementation of the VASG. \\ 
\indent The most attractive solution for the implementation of the VPSG is the geometric phase (\citealt{pancharatnam1956generalized}; \citealt{berry1987adiabatic}){, which applies the required phase to the opposite circular polarization states}. 
The geometric phase is introduced when the fast-axis angle of a half-wave retarder is spatially varying. 
The phase that is induced is twice the fast-axis angle, and is opposite for the opposite circular polarization states. 
Due to its geometric {origin}, the geometric phase is completely achromatic. 
However, the efficiency with which the phase is transferred to the light depends on retardance offsets from half wave. 
Increasing the retardance offset decreases the amount of light that acquires the desired phase. 
The VPSGs simulated in \autoref{sec:simulations} are implemented by geometric phase. \\
\indent Half-wave retarders with a spatially varying fast-axis angle can be constructed in various ways. 
For example, metamaterials have been used to induce geometric phase, but their efficiency is generally low \citep{mueller2017metasurface}.
The most mature and promising is liquid-crystal technology \citep{escuti2016controlling}.
By a direct-write system, the desired fast-axis angle can be printed into a liquid-crystal photo-alignment layer that that has been deposited on a substrate \citep{miskiewicz2014direct}. 
To achromatise the half-wave retarder, several layers of carefully designed, self-aligning birefringent liquid crystals can be deposited on top of the initial layer \citep{komanduri2013multi}. 
{In astronomy,  there have already been several successful (broadband) implementations of this technology: in coronagraphy (\citealt{mawet2009optical}; \citealt{snik2012vector}),  polarimetry (\citealt{tinyanont2018wirc+}; \citealt{snik2019snapshot}), wavefront sensing (\citealt{haffert2016generalised}; \citealt{doelman2019simultaneous}), and interferometry \citep{doelman2018multiplexed}.} \\
\indent {The major drawback of liquid-crystal technology is when there are retardance offsets from half-wave, as the efficiency with which the light accumulates the desired phase decreases.}
{The light that does not acquire the desired phase will form an on-axis PSF, which is regularly referred to as the leakage.}
{In coronagraphy, the leakage severely limits the coronagraphic performance of the liquid-crystal optic (\citealt{bos2018fully}; \citealt{doelman2020minimizing}).}
{However, for the VSG the impact is much less severe, because the leakage will overlap with the stellar PSF and therefore be occulted by the coronagraph.}
{The relative intensity of the VSG will be affected,
but this effect will be relatively small as $I_s \propto (1-L)$ \citep{bos2019focal}, with $L$ the leakage strength.} 
{The leakage strength is generally on the order of $\sim2\cdot10^{-2}$ \citep{doelman2017patterned} for broadband devices.}
\section{Discussion and conclusion}\label{sec:conclusions} 
Here, we show that by applying opposite modulation on opposite polarization states in the pupil-plane amplitude or phase, a speckle grid in the focal plane is generated that can be used as a photometric and astrometric reference.
We refer to this as the Vector Speckle Grid (VSG). 
In this implementation, the speckle grid will not interfere with the central stellar PSF and will therefore be effectively incoherent. 
This greatly decreases the photometric and astrometric errors when the PSF is distorted by aberrations. 
Furthermore, we identified that the remaining limiting factor is the cross-talk between the speckles in the grid itself. 
This can be mitigated by increasing the separation between the speckles. \\
\indent We performed simulations {with various atmospheric conditions} to quantify the performance increase with respect to regular SGs. 
{We find that for the conditions simulated, the VSGs improve the photometric and astrometric errors by a factor of $\sim$20 and $\sim$5, respectively, reaching a $\sim$0.3-0.8\% photometric and a $\sim$$3-10\cdot10^{-3}$ $\lambda/D$ astrometric error on short exposure images.}
We note that the performance increase depends on the brightness difference between the speckle and the residual {speckle} background. 
When the brightness difference increases, the performance increase is more moderate. 
If the speckles are dimmer, the performance increase is higher. 
We also investigated the effects of partially polarized light on the performance of the VSGs. 
The simulations showed that when the degree of polarization was below 5\%, the performance was barely affected. 
The {polarization signal} introduced by the telescope and instrument is on this level or less, and therefore not relevant.
We note that it is hard to predict what the on-sky performance will be as it is notoriously difficult to capture all relevant effects in simulation \citep{guyon2019technology}.
{Therefore, these results are an indication of the performance increase that the VSGs could bring.} 
{We also note that the performance of the ASG and PSG reported in these simulations is better than what has been reported on-sky, while the duration of the simulations is much shorter than the actual observations.}
{This is because these simulations only consider the effects of AO-corrected atmospheric wavefront errors, while observations are also affected by noise processes that generate background speckles with much longer decorrelation timescales.} 
{The VSG would also be incoherent to these background speckle noise sources.} \\
\indent We identified that the most attractive implementation of VSGs would be a Vector Phase Speckle Grid (VPSG) by the geometric phase. 
Liquid-crystal technology allows for a broadband half-wave retarder with a varying fast-axis angle that will induce the geometric phase on the light.
This has the major advantage that the VPSG can be implemented as a one pupil-plane optic. \\ 
\indent Implementing the VPSG by liquid-crystal technology has the following advantages: {it achieves instantaneous incoherency, it is a static component and therefore easy to calibrate, the artificial speckles can be positioned anywhere in the focal plane, the geometric phase is achromatic and therefore the speckles have a constant brightness with wavelength.}
Another advantage, not discussed in this paper, is that the VPSG could be multiplexed with holograms for wavefront sensing \citep{wilby2017coronagraphic}.
The main disadvantage of the VPSG is that the position of the speckles is fixed, making accurate background estimates more difficult{, and decreasing the flexibility of speckle grid positioning}. \\ 
\indent To conclude, the VSG has proven to be a promising method for generating speckle grids as photometric and astrometric references.
We show that the VSG reaches a satisfactory performance in simulation, and the next steps will be {an investigation of broadband effects,} a lab demonstration, and subsequent on-sky tests. 
The VSG could be part of the future upgrades of SPHERE and GPI (\citealt{beuzit2018possible}; \citealt{chilcote2018upgrading}).
\begin{acknowledgements} 
{The author thanks the referee for comments on the manuscript that improved this work.
The author also} thanks F. Snik for his comments on the manuscript.
The research of S.P. Bos leading to these results has received funding from the European Research Council under ERC Starting Grant agreement 678194 (FALCONER). 
This research made use of HCIPy, an open-source object-oriented framework written in Python for performing end-to-end simulations of high-contrast imaging instruments \citep{por2018hcipy}. 
This research used the following Python libraries: Scipy \citep{jones2014scipy}, Numpy \citep{walt2011numpy}, and Matplotlib \citep{Hunter:2007}.
\end{acknowledgements}

\bibliography{report} 
\bibliographystyle{aa} 

\end{document}